\documentclass[a4paper,12pt]{article}
\usepackage{epsfig}

\newcommand{\mrm}[1]{\mbox{\rm #1}}
\def\beq{\begin{equation}}
\def\eeq{\end{equation}}
\def\bea{\begin{eqnarray}}
\def\eea{\end{eqnarray}}
\def\bq{\begin{quote}}
\def\eq{\end{quote}}

\def\gappeq{\mathrel{\rlap {\raise.5ex\hbox{$>$}}
{\lower.5ex\hbox{$\sim$}}}}

\def\lappeq{\mathrel{\rlap{\raise.5ex\hbox{$<$}}
{\lower.5ex\hbox{$\sim$}}}}

\begin{document}
\pagestyle{empty}
\begin{flushright}
{CERN-TH/2003-073}\\
hep-ph/0303242\\
\end{flushright}
\vspace*{5mm}
\begin{center}
{\large {\bf Sneutrino Inflation in the Light of WMAP: \\
Reheating, Leptogenesis and Flavour-Violating Lepton Decays}} \\
\vspace*{1cm}
{\bf John~Ellis}$^1$, {\bf Martti~Raidal}$^{1,2}$ and {\bf 
T.~Yanagida}$^3$
\vspace{0.3cm}

$^1$ TH Division, CERN, CH-1211 Geneva 23, Switzerland \\
$^2$ National Institute of Chemical Physics and Biophysics,
Tallinn 10143, Estonia \\
$^3$ Department of Physics, University of Tokyo, Tokyo 113-0033, Japan \\
\vspace*{2cm}
{\bf ABSTRACT} \\ 
\end{center}
\vspace*{5mm}
\noindent

We reconsider the possibility that inflation was driven by a sneutrino -
the scalar supersymmetric partner of a heavy singlet neutrino - in the
minimal seesaw model of neutrino masses. We show that this model is
consistent with data on the cosmic microwave background (CMB), including
those from the WMAP satellite. We derive and implement the CMB constraints
on sneutrino properties, calculate reheating and the cosmological baryon
asymmetry arising via direct leptogenesis from 
sneutrino decays following sneutrino inflation,
and relate them to light neutrino masses.  We show that
this scenario is compatible with a low reheating temperature that avoids
the gravitino problem, and calculate its predictions for flavour-violating
decays of charged leptons. We find that $\mu \to e \gamma$ should occur
close to the present experimental upper limits, as might also $\tau \to
\mu \gamma$.

\vspace*{2cm}
\noindent

\begin{flushleft} CERN-TH/2003-073 \\
March 2003
\end{flushleft}
\vfill\eject
%\pagestyle{empty}
%\clearpage\mbox{}\clearpage

\setcounter{page}{1}
\pagestyle{plain}

%INSERT YOUR TEXT HERE

\section{Introduction}

Inflation~\cite{inf} has become the paradigm for early cosmology, particularly
following the recent spectacular CMB data from the WMAP satellite~\cite{wmap}, which
strengthen the case made for inflation by earlier data, by measuring an
almost scale-free spectrum of Gaussian adiabatic density fluctuations
exhibiting power and polarization on super-horizon scales, just as
predicted by simple field-theoretical models of inflation. As we review
below, the scale of the vacuum energy during inflation was apparently
$\sim 10^{16}$~GeV, comparable to the expected GUT scale, so CMB
measurements offer us a direct window on ultra-high-energy physics.

Ever since inflation was proposed, it has been a puzzle how to integrate 
it with ideas in particle physics. For example, a naive GUT Higgs field 
would give excessive density perturbations, and no convincing concrete 
string-theoretical model has yet emerged. In this conceptual vacuum, 
models based on simple singlet scalar fields have held sway~\cite{inf}. 
The simplest 
of these are chaotic inflation models based on exponential or power-law 
potentials, of which $\phi^4$ and $\phi^2$ are the only renormalizable 
examples. The WMAP collaboration has made so bold as to claim that such a 
$\phi^4$ model is 
excluded at the 3-$\sigma$ level~\footnote{This argument applies  
{\it a fortiori} to models with $\phi^{n > 4}$ potentials.}, a conclusion 
which would merit further support~\cite{barger,realbarger}. Nevertheless, 
it is clear that a 
$\phi^2$ model would be favoured.

We reconsider in this paper the possibility that the inflaton could in
fact be related to the other dramatic recent development in fundamental
physics, namely the discovery of neutrino masses~\cite{mnuexp}. The simplest models of 
neutrino masses invoke heavy singlet neutrinos that give masses to the 
light neutrinos via the seesaw mechanism~\cite{seesaw}. The heavy singlet neutrinos are 
usually postulated to weigh $10^{10}$ to $10^{15}$~GeV, embracing the 
range where the inflaton mass should lie, according to WMAP {\it et al}. 
In supersymmetric models, the heavy singlet neutrinos have scalar partners 
with similar masses, {\it sneutrinos}, whose properties are ideal for 
playing the inflaton role~\cite{sn1}. In this paper, we discuss the simplest scenario 
in which the lightest heavy singlet sneutrino drives inflation. This 
scenario constrains in interesting ways many of the 18 parameters of 
the minimal seesaw model for generating three non-zero light neutrino 
masses.

This minimal sneutrino inflationary scenario (i) yields a simple ${1 \over
2} m^2 \phi^2$ potential with no quartic terms, with (ii) masses $m$ lying
naturally in the inflationary ballpark. The resulting (iii) spectral index
$n_s$, (iv) the running of $n_s$ and (v)  the relative tensor strength $r$
are all compatible with the data from WMAP and other 
experiments~\cite{wmap}.
Moreover, fixing $m \sim 2 \times 10^{13}$~GeV as required by the observed
density perturbations (vi) is compatible with a low reheating temperature
of the Universe that evades the gravitino problem~\cite{gr}, (vii) realizes
leptogenesis~\cite{fy,sn2} in a calculable and viable way, (viii) constrains neutrino
model parameters, and (ix) makes testable predictions for the
flavour-violating decays of charged leptons.

The main features of our scenario are the following.
{\it First}, reheating of 
the Universe is now 
due to the neutrino Yukawa couplings, and therefore can be
related to light neutrino masses and mixings. 
{\it Secondly},  
the lepton asymmetry is created in direct 
sneutrino-inflaton decays~\cite{sn2}.
There is only one parameter describing the efficiency of
leptogenesis in this minimal sneutrino inflationary scenario in all
leptogenesis regimes - the reheating temperature of the Universe - to 
which
the other relevant parameters can be related.
This should be compared with the general thermal 
leptogenesis case~\cite{fy,pluemi,bbp,gnrrs} 
which has two additional independent parameters, 
namely the lightest heavy neutrino mass and width. 
{\it Thirdly},
imposing the requirement of successful leptogenesis, we calculate
branching ratios for $\mu\to e\gamma$ and $\tau\to\mu\gamma$~\cite{lfv}, 
and the CP-violating observables~\cite{cp} like the
electric dipole moments of the electron and muon~\cite{edm}. All these leptonic
observables, as well as leptogenesis, are related to the measured neutrino
masses via a parametrization with a random orthogonal matrix~\cite{ci}. We show
that, in the minimal scenario discussed here, successful leptogenesis
implies a prediction for $\mu\to e\gamma$ in a very narrow band within
about one order of magnitude of the present experimental bound, whilst 
$\tau\to \mu\gamma$ might be somewhat further away.

Other sneutrino inflationary scenarios could be considered. For example,
the inflaton might be one of the heavier singlet sneutrinos, or two or
more sneutrinos might contribute to inflation, or one might play a role as
a curvaton~\cite{curv}. These alternatives certainly merit consideration, though they
would in general be less predictive. We find it remarkable that the
simplest sneutrino inflationary scenario considered here works as well as
it does.

\section{Chaotic Sneutrino Inflation}

We start by reviewing chaotic inflation~\cite{inf} with a $V = {1 
\over 2} m^2 
\phi^2$ potential - the form expected for a heavy singlet sneutrino - in 
light of WMAP~\cite{wmap}. Defining $M_P \equiv 1/\sqrt{8 \pi G_N} 
\simeq 2.4 \times 10^{18}$~GeV, the conventional slow-roll inflationary 
parameters are
\beq
\epsilon \equiv {1 \over 2} M_P^2 \left( {V^\prime \over V} \right)^2 = 
{2 M_P^2 \over \phi_I^2}, \;
\eta \equiv M_P^2 \left( {V^{\prime\prime} \over V} \right) = {2 M_P^2 
\over \phi_I^2}, \;
\xi \equiv M_P^4 \left( {V V^{\prime\prime\prime} \over V^2} \right) = 
0,
\label{slowroll}
\eeq
where $\phi_I$ denotes the {\it a priori} unknown inflaton field value 
during inflation at a typical CMB scale $k$. The overall scale of the 
inflationary potential is normalized by the WMAP data on density 
fluctuations:
\beq
\Delta_R^2 = {V \over 24 \pi^2 M_P^2 \epsilon} = 2.95 \times 10^{-9} A ~~~:~~~
A = 0.77 \pm 0.07,
\label{normn}
\eeq
yielding
\beq
V^{1 \over 4} = M_P ^4\sqrt{\epsilon \times 24 \pi^2 \times 2.27 \times 
10^{-9}} = 0.027 M_P \times \epsilon^{1 \over 4},
\label{WMAP}
\eeq
corresponding to
\beq
m^{1 \over 2} \phi_I \; = \; 0.038 \times M_P^{3 \over 2}
\label{onecombn}
\eeq
in any simple chaotic $\phi^2$ inflationary model, such as the sneutrino 
model explore here. The number of e-foldings after the 
generation of the CMB density fluctuations observed by COBE is estimated 
to be
\beq
N_{COBE} \; = 62 - {\rm ln} \left( {10^{16}~{\rm GeV} \over V_{end}^{1/4}} 
\right) - {1 \over 3} {\rm ln} \left( { V_{end}^{1/4} \over \rho_{RH} } 
\right),
\label{NCOBE}
\eeq
where $\rho_{RH}$ is the energy density of the Universe when it is 
reheated after inflation. The second term in (\ref{NCOBE}) is negligible in 
our model, whereas the third term could 
be as large as $(-8)$ for a reheating temperature $T_{RH}$ as low as 
$10^6$~GeV. Conservatively, we take
$N \simeq 50$. In a $\phi^2$ inflationary model, this implies
\beq
N \; = {1 \over 4} {\phi^2_I \over M_P^2} \simeq \; 50,
\label{fixN}
\eeq
corresponding to
\beq
\phi^2_I \simeq 200 \times M_P^2.
\label{fixphi}
\eeq
Inserting this requirement into the WMAP normalization condition 
(\ref{WMAP}), we find the following required mass for any quadratic 
inflaton:
\beq
m \; \simeq 1.8 \times 10^{13}~{\rm GeV}.
\label{phimass}
\eeq
As already mentioned, this is comfortably within the range of heavy 
singlet (s)neutrino masses usually considered, namely $m_N \sim 10^{10}$ 
to $10^{15}$~GeV.

Is this simple $\phi^2$ sneutrino model compatible with the WMAP data? The 
primary CMB observables are the spectral index
\beq
n_s = 1 - 6 \epsilon + 2 \eta = 1 - {8 M_P^2 \over \phi^2_I} \simeq 
0.96,
\label{ns}
\eeq
the tensor-to scalar ratio
\beq
r \equiv {A_T \over A_S} = 16 \epsilon = {32 M_P^2 \over \phi^2_I} \simeq 
0.16,
\label{r}
\eeq
and the spectral-index running
\beq
{d n_s \over d {\rm ln} k} = {2 \over 3} \left[ \left( n_s - 1 \right)^2 - 
4 \eta^2 \right] + 2 \xi =   {32 M_P^4 \over \phi^4_I}  \simeq  8 \times 10^{-4}.
\label{values}
\eeq
The value of $n_s$ extracted from WMAP data depends whether, for example,
one combines them with other CMB and/or large-scale structure data. 
However, the $\phi^2$ sneutrino model value $n_s \simeq 0.96$ appears to 
be compatible with the data at the 1-$\sigma$ level. The $\phi^2$ sneutrino 
model value $r\simeq 0.16$ for the relative tensor strength is also 
compatible with the WMAP data. One of the most interesting features of the 
WMAP analysis is the possibility that ${d n_s / d {\rm ln} k}$ might 
differ from zero. The $\phi^2$ sneutrino model value ${d n_s / d {\rm ln} 
k} \simeq 8 \times 10^{-4}$ derived above is negligible compared with the 
WMAP preferred value and its uncertainties. However, ${d n_s / d {\rm ln} 
k} = 0$ appears to be compatible with the WMAP analysis at the 2-$\sigma$ 
level or better, so we do not regard this as a death-knell for the 
$\phi^2$ sneutrino model~\footnote{In fact, we note that the favoured 
individual values for $n_s, r$ and ${d n_s / d {\rm ln}
k}$ reported in an independent analysis~\cite{realbarger} {\it all 
coincide} with the 
$\phi^2$ sneutrino model values, within the latter's errors!}.

\section{Reheating and Leptogenesis}

Before addressing leptogenesis in this sneutrino model for inflation in all
calculational details, we 
first comment on the reheating temperature $T_{RH}$ following the 
inflationary epoch. Assuming, as usual, that the sneutrino inflaton decays 
when the the Hubble expansion rate $H \sim m$, and that the expansion rate 
of the Universe is then dominated effectively by non-relativistic matter 
until $H \sim \Gamma_\phi$,  where $\Gamma_\phi$ is the inflaton decay width, 
we estimate
\beq
T_{RH} = \left( 
\frac{90}{ \pi^2 g_*} 
\right)^{1 \over 4} \sqrt{\Gamma_\phi M_P} ,
\label{TRH}
\eeq
where $g_*$ is the number of effective relativistic degrees of freedom in 
the reheated Universe.
In the minimal sneutrino inflation scenario considered here we have
$\phi\equiv \tilde N_1,$ $m \equiv M_{N_1}$ and 
\bea
\Gamma_\phi\equiv \Gamma_{N_1} = 
\frac{1}{4 \pi} (Y_\nu Y_\nu^\dagger)_{11} M_{N_1},
\eea
where $Y_\nu$ is the neutrino Dirac Yukawa matrix.
If the relevant neutrino Yukawa coupling 
$(Y_\nu Y_\nu^\dagger)_{11} \sim 1$, the previous choice $m=M_{N_1} \simeq 
2 \times 10^{13}$~GeV would yield $T_{RH} > 10^{14}$~GeV, considerably greater 
than $m$ itself~\footnote{Even such a large value of $(Y_\nu Y_\nu^\dagger)_{11}$ 
would not alter significantly the $\phi^2$ sneutrino model 
prediction for ${d n_s / d {\rm ln} k}$.}.
Such a large value of
$T_{RH}$ would be {\it very problematic} for the thermal production of
gravitinos~\cite{gr}. However, it is certainly possible that $(Y_\nu
Y_\nu^\dagger)_{11}\ll 1$, in which case $T_{RH}$ could be much lower, as
we discuss in more detail below. Alternatively, one may consider more
complicated scenarios, in which three sneutrino species may share the
inflaton and/or curvaton roles between them.

We now present more details of reheating and leptogenesis.
In general, inflaton decay and the reheating of the Universe are described 
by the following set of Boltzmann equations~\cite{gkr}
\bea
\frac{\mrm{d}\rho_\phi}{\mrm{d} t} & = &
- 3 H \rho_\phi - \Gamma_\phi \rho_\phi , \nonumber \\
\frac{\mrm{d}\rho_R}{\mrm{d} t} & = &
- 4  H \rho_R +\Gamma_\phi \rho_\phi , 
\label{rehboltz} \\
H & = & \frac{\mrm{d} R}{R \mrm{d} t}=\sqrt{8 \pi G_N (\rho_\phi + \rho_R)/3},
\label{H}
\eea
where
$\rho_\phi$ is the energy density of the inflaton field,  
$\rho_R$ describes the energy density of the thermalized decay products
and essentially defines the temperature via
\bea
\rho_R = \frac{\pi^2}{30} g_* T^4,  
\eea
$H$ is the Hubble constant and $G_N$ is the Newton constant.
Thus reheating can be described by two parameters,
the reheating temperature (\ref{TRH}), which is the highest temperature
of thermal plasma immediately {\it after} reheating is completed,
and the initial energy density of the 
inflaton field
\bea
\rho_\phi\simeq \frac{\pi^2 g_* T^8}{5 T_{RH}^4},
\eea
which determines the maximal plasma temperature
in the beginning of the reheating process.
In the following we use the parameter
\bea
z=\frac{M_{N_1}}{T}
\eea
to parametrize temperature.

The set of Boltzmann equations describing 
the inflaton decay and reheating, the
creation and decays of thermal heavy neutrinos and sneutrinos, and 
the generation of a lepton asymmetry, is given by
\bea
Z \frac{\mrm{d}\rho_\phi}{\mrm{d}z} & = &
-\frac{3 \rho_\phi}{z} - \frac{\Gamma_\phi \rho_\phi}{zH} , 
\label{b1} \\ 
HZz \frac{\mrm{d} Y_{N_1}}{\mrm{d}z} & = &
-\frac{3\Gamma_\phi \rho_\phi}{4 \rho_R} Y_{N_1}  - \frac{1}{s} (remaining)  
\label{b2} \\
HZz \frac{\mrm{d} Y_{\tilde N_+}}{\mrm{d}z} & = &
-\frac{3\Gamma_\phi \rho_\phi}{4 \rho_R} Y_{\tilde N_+}  - \frac{1}{s} (remaining) 
\label{b3}  \\
HZz \frac{\mrm{d} Y_{\tilde N_-}}{\mrm{d}z} & = &
-\frac{3\Gamma_\phi \rho_\phi}{4 \rho_R} Y_{\tilde N_-}  - \frac{1}{s} (remaining)  
\label{b4} \\
HZz \frac{\mrm{d} Y_{L_f}}{\mrm{d}z} & = &
-\frac{3\Gamma_\phi \rho_\phi}{4 \rho_R} Y_{L_f} 
+\frac{\Gamma_\phi \rho_\phi}{2 s M_{N_1}} \epsilon_1 - \frac{1}{s} (remaining) 
\label{b5} \\
HZz \frac{\mrm{d} Y_{L_s}}{\mrm{d}z} & = &
-\frac{3\Gamma_\phi \rho_\phi}{4 \rho_R} Y_{L_s} 
+\frac{\Gamma_\phi \rho_\phi}{2 s M_{N_1}} \epsilon_1 - \frac{1}{s} (remaining) 
\label{b6} \\
H & = & \sqrt{8 \pi G_N (\rho_\phi + \rho_R)/3} ,
\label{b7} 
\eea
where
\bea
Z \equiv 1-\frac{\Gamma_\phi \rho_\phi}{4 H \rho_R},
\eea
$\tilde N_\pm \equiv \tilde N_1 \pm \tilde N_1^\dagger$, and $Y_{N_1},$ 
$Y_{\tilde N_\pm},$
$Y_{L_f},$ $Y_{L_s},$ denote the number-density-to-entropy ratios, 
$Y=n/s$, for the heavy neutrinos, sneutrinos and lepton asymmetries in 
fermions and scalars, respectively.
The terms denoted by $remaining$ are the usual ones for thermal 
leptogenesis, and
can be obtained from~\cite{pluemi} by using $H(M_{N_1})=z^2 H$. 
We do not write out their lengthy expressions in full here.
The first terms on the r.h.s. of (\ref{b2}-\ref{b6}) are the dilution
factors of $Y=n/s$ due to entropy production in the inflaton
$\phi\equiv \tilde N_1$ decays described by (\ref{b1}). The second terms 
on the r.h.s. of (\ref{b5}), (\ref{b6}) describe lepton
asymmetry generation in the decays of the coherent inflaton field.
Identifying and studying 
the parameter space in which leptogenesis is predominantly
direct is one of the aims of this paper.

We are now ready to study (\ref{b1}-\ref{b7}).
First we work out general results on reheating and leptogenesis 
in the sneutrino inflation scenario, allowing $M_{N_1}$ to vary as a free 
parameter.
In this case, the reheating and leptogenesis efficiency is described
by two parameters, namely $M_{N_1}$ and a parameter describing the decays 
of the sneutrino inflaton. This can be chosen to be either 
$\tilde m_1= (Y_\nu Y_\nu^\dagger)_{11} v^2 \sin^2\beta/M_{N_1}$ or, 
more appropriately for this scenario, the reheating temperature of
the Universe $T_{RH}$ given by (\ref{TRH}).
For the CP asymmetry in (s)neutrino decays, we take the maximal value
for hierarchical light neutrinos, given by~\cite{di2}:
\bea
|\epsilon_1^{max}(M_{N_1})|=\frac{3}{8 \pi} 
\frac{M_{N_1} \sqrt{\Delta m^2_{atm}}}{v^2 \sin^2\beta}.
\eea
This choice allows us to study the minimal values for $M_{N_1}$ and 
$T_{RH}$ allowed by leptogenesis. Later, we will focus our attention on 
exact values of $\epsilon_1$~\cite{vissani}.

\begin{figure}[t]
\centerline{\epsfxsize = 0.5\textwidth \epsffile{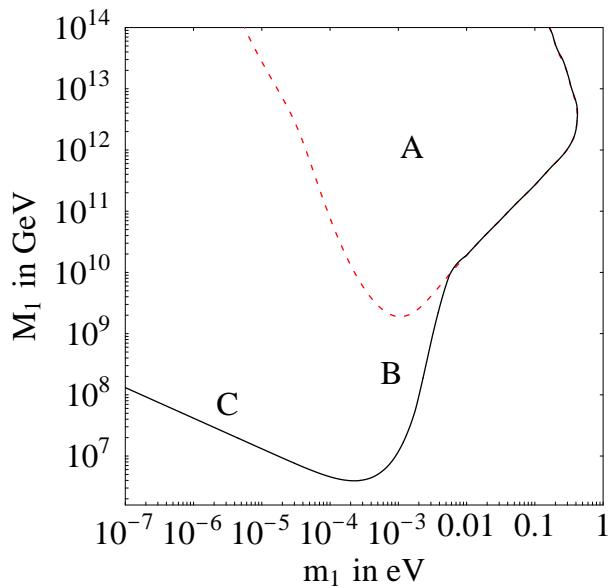}
%\hfill \epsfxsize = 0.5\textwidth \epsffile{effm1sn.eps}
}
\caption{\it
Lower bound (solid curve) on $M_{N_1}$ as a function of  $\tilde 
m_1$ for $Y_B > 7.8\times 10^{-11}$, assuming a maximal CP asymmetry 
$\epsilon_1^{max}(M_{N_1})$. Successful leptogenesis is possible in the 
area above the solid curve. In the area bounded by the red dashed curve, 
leptogenesis is entirely thermal. 
\vspace*{0.5cm}}
\label{fig1}
\end{figure}

\begin{figure}[t]
\centerline{\epsfxsize = 0.5\textwidth \epsffile{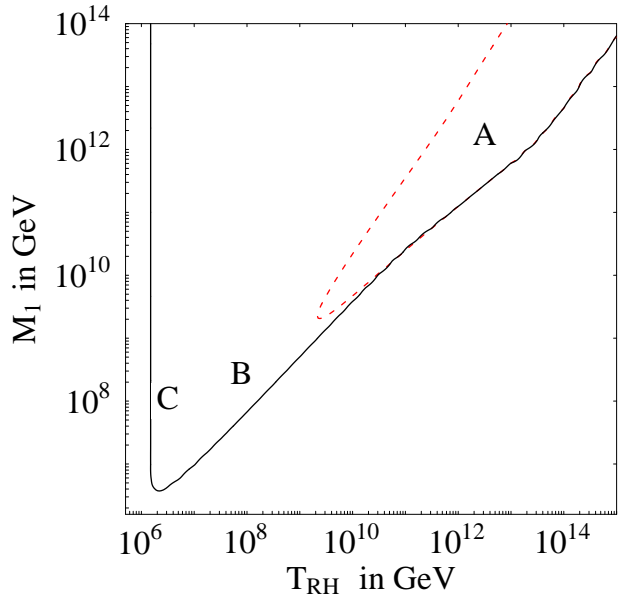}   
%\hfill \epsfxsize = 0.5\textwidth \epsffile{efftrsn.eps}
}
\caption{\it
The solid curve bounds the region allowed for leptogenesis in the
$(T_{RH},\,M_{N_1})$ plane, again obtained assuming $Y_B > 7.8\times
10^{-11}$ and the maximal CP asymmetry $\epsilon_1^{max}(M_{N_1}).$ In the
area bounded by the red dashed curve leptogenesis is entirely thermal.
\vspace*{0.5cm}}
\label{fig2}
\end{figure}

Solutions to (\ref{b1}-\ref{b7}) are presented in Figs.~\ref{fig1} and 
\ref{fig2}. We plot in Fig.~\ref{fig1} the
parameter space in the $(M_{N_1},\,\tilde m_{1})$ plane that leads to
successful leptogenesis.  This parameter space has three distinctive parts
with very different physics.

In the area bounded by the red dashed curve, denoted by A, leptogenesis is
entirely thermal. 
This region has been studied in detail in~\cite{gnrrs}.
Whatever lepton asymmetry is generated initially in the
decay of the sneutrino inflaton is washed out by thermal effects, and the
observed baryon asymmetry is generated by the out-of-equilibrium decays of
thermally created singlet neutrinos and sneutrinos. As seen in
Fig.~\ref{fig2}, in our scenario 
this parameter space corresponds to high $M_{N_1}$ and
high $T_{RH}$ values.

The area B below the dashed curve and extending down to the minimum value
$M_{N_1}=4\times 10^6$ GeV in Fig.~\ref{fig1} is the region of parameter
space where there is a delicate cancellation between direct lepton
asymmetry production in sneutrino inflaton decays and thermal washout.
This region cannot be studied without solving the Boltzmann equations
numerically. However, it roughly corresponds to $T_{RH}\sim M_{N_1}$ as
seen in Fig.~\ref{fig2}.

The area denoted by C has $T_{RH}\ll M_{N_1}$. Since the maximal CP
asymmetry scales with $M_{N_1},$ the line presented corresponds to a
constant reheating temperature. 
Notice that in Fig.~\ref{fig1} this line is terminated at $\tilde m_1=10^{-7}$.
As seen in Fig.~\ref{fig2}, it continues linearly to high values of $M_{N_1}$.
In this area, leptogenesis is entirely
given by the decays of cold sneutrino inflatons, a scenario studied 
previously in~\cite{sn2}. 
In this case the details of reheating are not important for
our analyses.  
To calculate the lepton asymmetry to entropy density
ratio $Y_L=n_L/s$ in inflaton decays we need to know the
produced entropy density
\bea
s=\frac{2\pi^2}{45} g_* T_{RH}^3,
\eea
and to take into account that inflaton dominates the
Universe. In this case one obtains~\cite{sn2}
\bea
Y_L=\frac{3}{4} \epsilon_1 \frac{T_{RH}}{M_{N_1}},
\eea
where $\epsilon_1$ is the CP asymmetry in 
$\phi\equiv \tilde N_1$ decays.
The observed baryon asymmetry of the
Universe gives a lower bound on
the reheating temperature $T_{RH}>10^6$ GeV.

We consider now the most constrained scenario in which the inflaton is the
lightest sneutrino, which requires $M_{N_3} > M_{N_2} > M_{N_1} \simeq 2 \times
10^{13}$~GeV. This implies that our problem is completely characterized by 
 only one parameter, either $\tilde m_1$ or $T_{RH}$.
As we see in both Figs.~\ref{fig1} and \ref{fig2}, the line for  
$M_{N_1} \simeq 2 \times 10^{13}$~GeV
traverses both the regions A, and C, the former corresponding to high
$T_{RH}$, as seen in Fig.~\ref{fig2}. However, $T_{RH}$ may also be 
low even in the minimal seesaw model, as seen in Fig.~\ref{fig2}.

The cosmological gravitino problem suggests that $T_{RH} \lappeq 10^8$ GeV
might be the most interesting, which would correspond to very small $\tilde m_1$,  
far away from the thermal region A and deep in the
region C where leptogenesis arises from the direct decays of cold
sneutrinos. We concentrate on this option here.
This limit requires very small Yukawa couplings $( Y_\nu 
Y_\nu^\dagger )_{11} \lappeq 10^{-12}$, whilst other Yukawa couplings can 
be ${\cal O} (1)$. This possibility may be made natural, e.g., by 
postulating a $Z_2$ matter parity under which only $N_1$ is odd. In this 
case, the relevant Yukawa couplings $(Y_\nu)^1_j$ all vanish, but a 
Majorana mass for $N_1$ is still allowed. A more sophisticated model 
postulates a $Z_7$ discrete family symmetry with charges $Y_{FN} = (4, 0, 
0)$ for the $N_i$, $(2, 1, 1)$ for the ${\mathbf {\bar 5}}$ 
representations of SU(5), and $(2, 1, 0)$ for the ${\mathbf 10}$ 
representations of SU(5). Assuming a gauge-singlet field $\Phi$ with 
$Y_{FN} = -1$ and $\langle \Phi \rangle \equiv \epsilon$, we find $M_i = 
{\cal O}(\epsilon, 1, 1)$ and  $(Y_\nu)^1_j = {\cal O}(\epsilon^6, 
\epsilon^5, \epsilon^5)$, whilst the other Yukawa couplings are ${\cal 
O}(1)$, ${\cal O}(\epsilon)$ or ${\cal O}(\epsilon^2)$. If $\epsilon 
\simeq 1/17$, the $(Y_\nu)^1_j$ are sufficiently small for our purposes, 
whilst the quark and lepton mass matrices are of desirable form. 
Doubtless, one could construct better models with more effort, but this 
example serves as an existence proof for a low value of $T_{RH}$ in our 
scenario.

\section{Leptogenesis Predictions for Lepton Flavour Violation}

In this Section, we relate the results of the previous section on
direct leptogenesis to light neutrino masses, and make predictions on the
lepton-flavour-violating (LFV) decays. 
Thermal leptogenesis in this context  has been extensively
studied recently~\cite{lepto,er,fgy}. We first calculate neutrino Yukawa
couplings using the parametrization in terms of the light and heavy
neutrino masses, mixings and the orthogonal parameter matrix given
in~\cite{ci}. This allows us to calculate exactly the baryon asymmetry of
the Universe, since we know the CP asymmetry $\epsilon_1$ and the
reheating temperature of the Universe $T_{RH}.$ For neutrino parameters
yielding successful leptogenesis, we calculate the branching ratios of LFV
decays.

\begin{figure}[t]
\centerline{\epsfxsize = 0.5\textwidth \epsffile{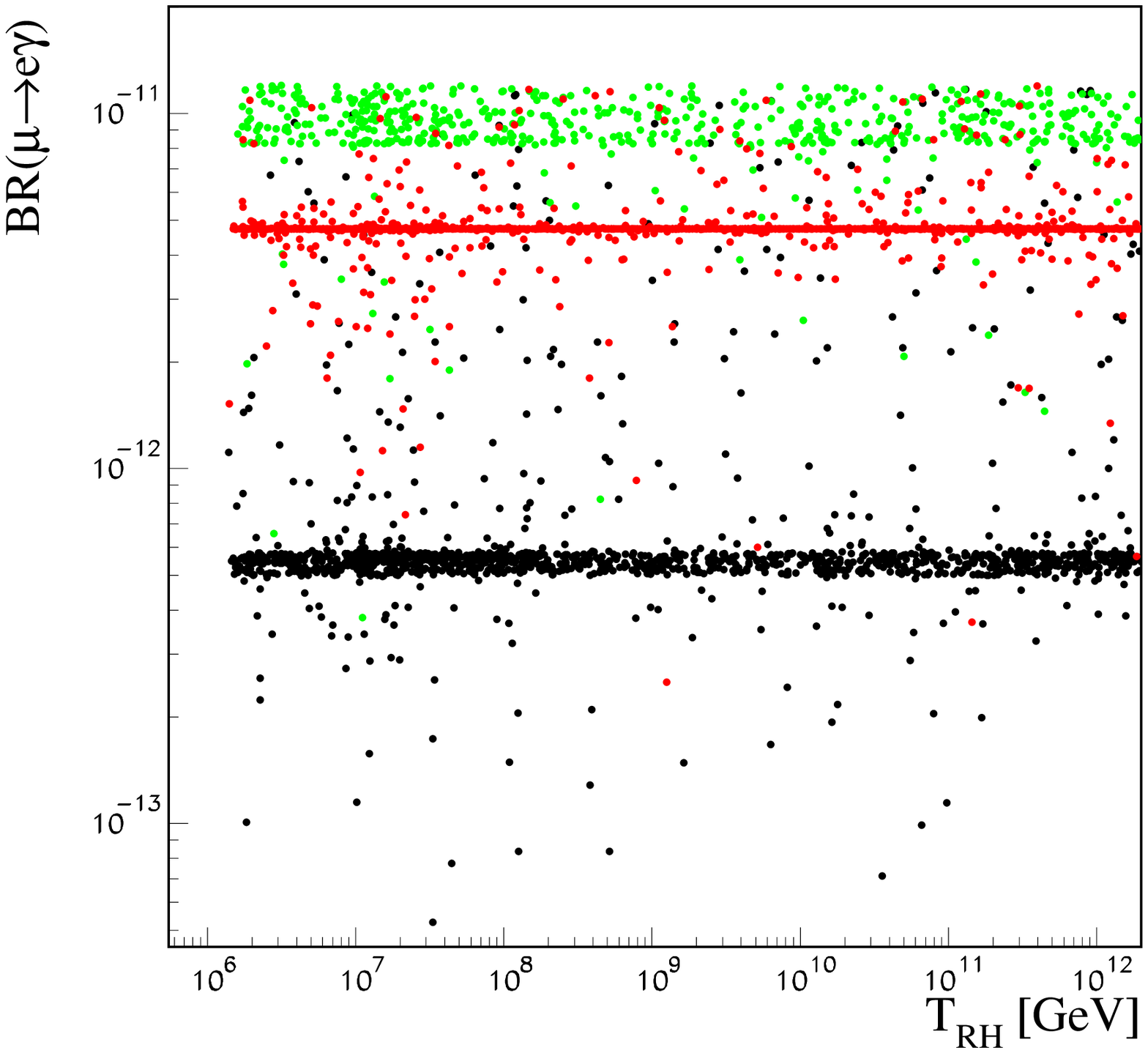}
\hfill \epsfxsize = 0.5\textwidth \epsffile{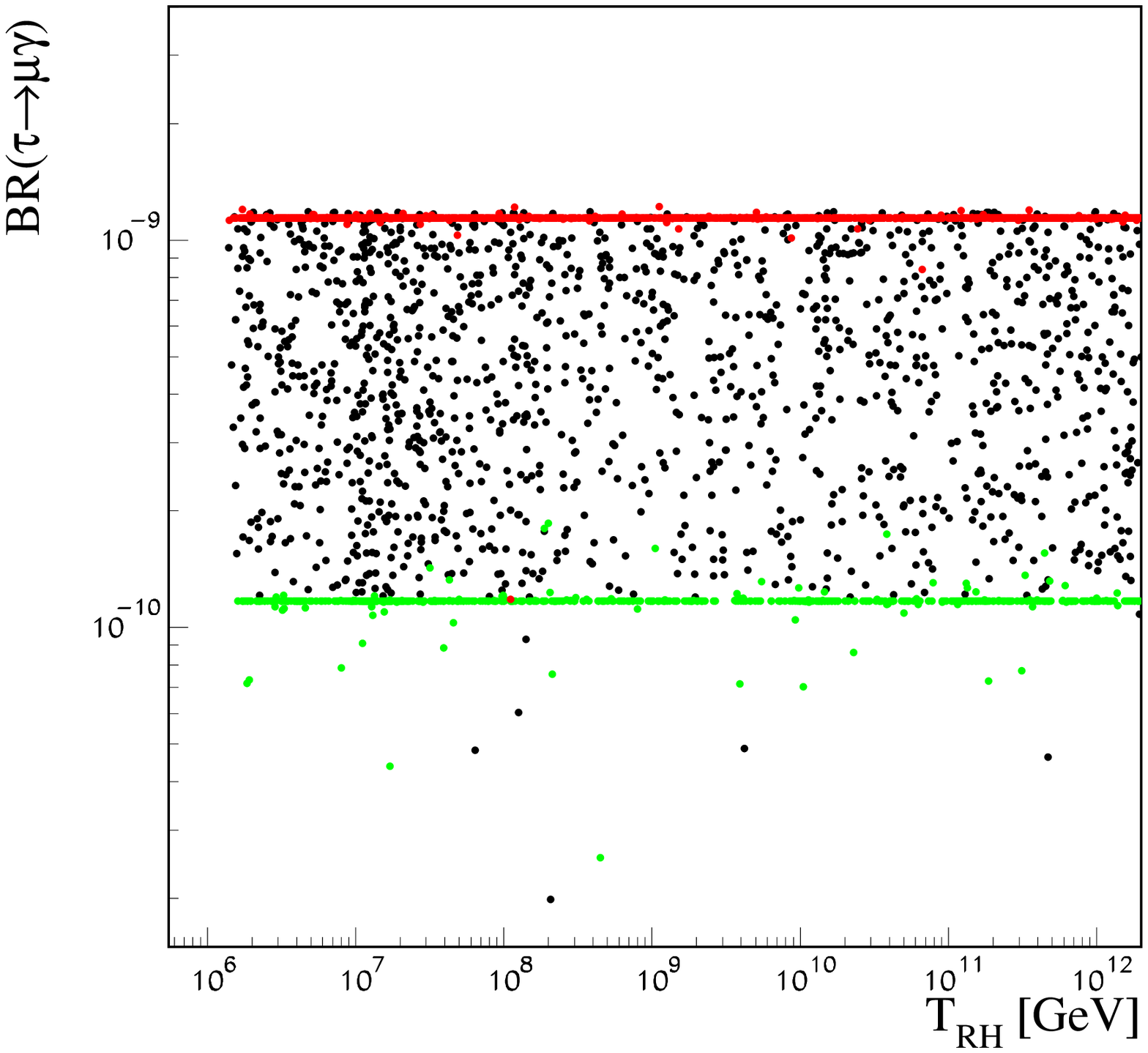}
}
\caption{\it
Calculations of BR$(\mu \to e \gamma)$ and BR$(\tau \to \mu \gamma)$
on left and right panels, respectively. Black points correspond to
$\sin \theta_{13} = 0.0$,  $M_2 = 10^{14}$~GeV, 
and $5 \times 10^{14}$~GeV $< M_3 < 5 \times 10^{15}$~GeV.
Red points correspond to $\sin \theta_{13} = 0.0$,  $M_2 = 5 \times 10^{14}$~GeV, 
and  $M_3 = 5 \times 10^{15}$~GeV, while green points correspond to
 $\sin \theta_{13} = 0.1$, 
$M_2 = 10^{14}$~GeV, and $M_3  = 5 \times 10^{14}$~GeV.
\vspace*{0.5cm}}
\label{fig3}
\end{figure}

There are 18 free parameters in the minimal seesaw model with three
non-zero light neutrinos, which we treat as follows. In making
Fig.~\ref{fig3}, we have taken the values of $\theta_{12}, \theta_{23}$,
$\Delta m^2_{12}$ and $\Delta m^2_{23}$ from neutrino oscillation
experiments. We randomly generate the lightest neutrino mass in the range
$0 < m_1 < 0.01$ eV and values of $\theta_{13}$ in the range $0 <
\theta_{13} < 0.1$ allowed by the Chooz experiment~\cite{chooz}, as we 
discuss later in
more detail. Motivated by our previous discussion of chaotic sneutrino
inflation, we fix the lightest heavy singlet sneutrino mass to be 
$M_1 = 2 \times 10^{13}$~GeV, and choose the following values of the heavier
singlet sneutrino masses: $M_2 = 10^{14}$~GeV or $M_2 =5 \times 10^{14}$~GeV, 
and $M_3$ in the range $5 \times 10^{14}$ to $5 \times
10^{15}$~GeV, as we also discuss later in more detail. This accounts for
nine of the 18 seesaw parameters.

The remaining 9 parameters are all generated randomly. These include the
three light-neutrino phases - the Maki-Nakagawa-Sakata oscillation phase
and the two Majorana phases.  Specification of the neutrino Yukawa
coupling matrix requires three more mixing angles and three more
CP-violating phases that are relevant to leptogenesis, in principle. The
plots in Fig.~\ref{fig3} are made by sampling randomly these nine
parameters. We apply one constraint, namely that the generated baryon
density falls within the $3-\sigma$ range required by cosmological
measurements, of which the most precise is now that by WMAP: $7.8\times
10^{-11} < Y_B < 1.0 \times 10^{-10}$~\cite{wmap}.

Making predictions for LFV decays also requires some hypotheses on the
parameters of the MSSM. We assume that the soft supersymmetry-breaking
mass parameters $m_0$ of the squarks and sleptons are universal, and
likewise the gaugino masses $m_{1/2}$, and we set the trilinear soft
supersymmetry-breaking parameter $A_0 = 0$ at the GUT scale. Motivated by
$g_\mu - 2$, we assume that the higgsino mixing parameter $\mu > 0$, and
choose the representative value $\tan \beta = 10$. We take into account
laboratory and cosmological constraints on the MSSM, including limits on
the relic density of cold dark matter. WMAP provides the most stringent
bound on the latter, which we assume to be dominated by the lightest
neutralino $\chi$: $0.094 < \Omega_\chi h^2 < 0.129$. For $\tan \beta =
10$, the allowed domain of the $(m_{1/2}, m_0)$ plane is an almost
linear strip extending from $(m_{1/2}, m_0) = (300, 70)$~GeV to $(900,
200)$~GeV~\cite{dm}. For illustrative purposes, we choose $(m_{1/2}, m_0) = (800,
170)$~GeV and comment later on the variation with $m_{1/2}$.

Panel (a) of Fig.~\ref{fig3} presents results on the branching ratio BR
for $\mu \to e \gamma$ decay. We see immediately that values of $T_{RH}$
anywhere between $2 \times 10^6$~GeV and $ 10^{12}$~GeV are
attainable in principle. The lower bound is due to the lower
bound on the CP asymmetry, while the upper bound comes
from the gravitino problem. 
The black points in panel (a)  correspond to the choice 
$\sin \theta_{13} = 0.0$,  $M_2 = 10^{14}$~GeV, 
and $5 \times 10^{14}$~GeV $< M_3 < 5 \times 10^{15}$~GeV.
The red points correspond to $\sin \theta_{13} = 0.0$,  $M_2 = 5 \times 10^{14}$~GeV, 
and  $M_3 = 5 \times 10^{15}$~GeV, while the green points correspond to
$\sin \theta_{13} = 0.1$, $M_2 = 10^{14}$~GeV, and $M_3  = 5 \times 10^{14}$~GeV.
We see a very striking narrow, densely populated  bands for BR$(\mu \to e \gamma)$, 
with some outlying points at both larger and smaller
values of BR$(\mu \to e \gamma)$. 
The width of the black band is due to variation of $M_{N_3}$ showing that  
BR$(\mu \to e \gamma)$ is not very sensitive to it. However, 
BR$(\mu \to e \gamma)$ strongly depends on $M_{N_2}$ and  $\sin \theta_{13}$
as seen by the red and green points, respectively. Since
BR$(\mu \to e \gamma)$ scales approximately as $m_{1/2}^{-4}$, the lower
strip for $\sin \theta_{13} = 0$ would move up close to the experimental
limit if $m_{1/2} \sim 500$~GeV, and the upper strip for $\sin \theta_{13}
= 0.1$ would be excluded by experiment.

Panel (b) of Fig.~\ref{fig3} presents the corresponding results for
BR$(\tau \to \mu \gamma)$ with the same colour code for the parameters. 
This figure shows that BR$(\tau \to \mu \gamma)$ depends strongly on 
 $M_{N_3}$, while the dependence on $\sin \theta_{13}$ 
and  on $M_{N_2}$ is negligible. 
The numerical values of BR$(\tau \to \mu \gamma)$
 are somewhat below the present experimental upper
limit BR$(\tau \to \mu \gamma) \sim 10^{-7}$, but we note that the results
would all be increased by an order of magnitude if $m_{1/2} \sim 500$~GeV.
In this case, panel (a) of Fig.~\ref{fig3} tells us that the experimental
bound on BR$(\mu \to e \gamma)$ would enforce $\sin \theta_{13} \ll 0.1$,
but this would still be compatible with BR$(\tau \to \mu \gamma) >
10^{-8}$.

As a result, Fig.~\ref{fig3} strongly suggests that fixing the observed
baryon asymmetry of the Universe for the direct sneutrino leptogenesis
($T_{RH}<2\times 10^{12}$~GeV $<M_{N_1}$) implies a prediction for the LFV
decays provided $M_{N_2}$ and/or $M_{N_3}$ are also fixed. This
observation can be understood in the case of hierarchical light and heavy
neutrino masses. Consider first $\mu \to e \gamma$ for $\sin
\theta_{13}=0$. It turns out that the $N_2$ couplings dominate in $(Y_\nu
Y_\nu^\dagger)_{21}$ which determines BR$(\mu \to e \gamma)$. Also, the
$M_{N_2}$ term dominates in $\epsilon_1$ which implies $Y_B \sim (Y_\nu
Y_\nu^\dagger)_{21}/\sqrt{(Y_\nu Y_\nu^\dagger)_{11}}$, because
cancellations among the phases are unnatural. In the parametrization with
the orthogonal matrix $R$, this implies $Y_B \sim R_{23}/R_{22}$. If fine
tunings are not allowed, the requirement $T_{RH}<M_{N_1}$ fixes
$R_{23}/R_{22}$ and therefore relates $Y_B$ to $\mu \to e \gamma$. For
more general cases, the behaviour of BR$(\mu \to e \gamma)$ is more
complicated and additional contributions occur. However, those new
contributions tend to enhance BR$(\mu \to e \gamma)$, as exemplified in
Fig.~\ref{fig3} by the green dots.

The behaviour of BR$(\tau \to \mu \gamma)$ is simpler. To leading order in
the largest parameters, $\tau \to \mu \gamma$ depends on the $N_3$
couplings and mass, leading to  
$(Y_\nu Y_\nu^\dagger)_{32}\sim (Y_\nu)^2_{33} U_{33}U_{23}^\dagger$,
independently of leptogenesis results.

We have to stress here that such definite predictions for LFV processes
can always be avoided by fine tuning the neutrino parameters, as seen by
several scattered points in Fig.~\ref{fig3}. Points with small BR$(\mu \to
e \gamma)$ can be systematically generated using the parametrization of
$Y_\nu$ by a Hermitian matrix~\cite{di}, and the predictions for the LFV
decays thereby washed away. However, in this case, the $M_{N_i}$ are
outputs of the parametrization, and cannot be fixed as required by the
present analyses of sneutrino inflation. Therefore the
parametrization~\cite{di} is not appropriate for our leptogenesis
scenario. Finally, we comment that such fine tunings are impossible in
simple models of neutrino masses~\cite{fgy}.

Another possibility for avoiding the LFV predictions is to allow the heavy
neutrinos to be partially degenerate in mass, which enhances 
the CP asymmetries~\cite{p}. In supersymmetric models, this possibility 
was considered in~\cite{ery}.

In addition to the quantities shown in Fig.~\ref{fig3}, we have also
examined BR$(\tau \to e \gamma)$, which is always far below the present
experimental bound BR$(\tau \to e \gamma) \sim 10^{-7}$, and the electron
and muon electric dipole moments. We find that $d_e < 10^{-33}$~e~cm, in
general, putting it beyond the foreseeable experimental reach, and $|
d_\mu / d_e | \sim m_\mu / m_e$, rendering $d_\mu$ also unobservably
small.

\section{Alternative Scenarios and Conclusions}

We have considered in this paper the simplest sneutrino inflation
scenario, in which the inflaton $\phi$ is identified with the lightest
sneutrino, and its decays are directly responsible for leptogenesis. 
We find it remarkable that this simple scenario is not already ruled out, and
have noted the strong constraints it must satisfy enable it to make strong
predictions, both for CMB observables and LFV decays. These might soon be
found or invalidated. In the latter case the motivation to study more complicated
sneutrino inflation scenarios would be increased.

$\bullet$
One possibility is that inflation might have been driven by a different 
sneutrino, not the lightest one. In this case, the lightest sneutrino 
could in principle be considerably lighter than the $2 \times 10^{13}$~GeV 
required for the inflaton. This would seem to make more plausible a low 
reheating temperature, as suggested by the gravitino problem. However, 
this problem is not necessarily a critical issue, as it can already be 
avoided in the simplest sneutrino inflation scenario, as we have seen. On the 
other hand, if the lightest sneutrino is not the inflaton, leptogenesis 
decouples from inflationary reheating,  and predictivity is diminished.

$\bullet$
A second possibility is that two or more sneutrinos contribute to
inflation. In this case, the model predictions for the CMB observables and
the sneutrino mass would in general be changed. 

$\bullet$
A related third possibility is that one or more sneutrinos might function
as a curvaton, which would also weaken the CMB and sneutrino mass
predictions.

For the moment, we do not see the need to adopt any of these more 
complicated scenarios, but they certainly merit investigation, even ahead 
of the probable demise of the simplest sneutrino inflation scenario 
investigated here.

\vskip 0.5in
\vbox{
\noindent{ {\bf Acknowledgements} } \\
\noindent  
We thank G. Giudice, M. Pl\"umacher, M. Postma, A. Riotto and A. Strumia  
for collaboration and discussions related to this work.
T.Y. thanks the members of the theory group in Rome I University for 
their hospitality. M.R. is partially supported by EU TMR
contract No.  HPMF-CT-2000-00460 and by ESF grant No. 5135,
and T.Y. by Grant-in-Aid for Scientific Research (S) 14102004.
}

\end{document}